 \documentclass[preprint2]{aastex}

\shorttitle{SED of the Crab nebula}
\shortauthors{Macias-Perez et al.}

\begin{document}


\title{Global spectral energy distribution of the Crab Nebula in the prospect of the Planck satellite polarisation calibration}


\author{J.~F.~Mac\'{\i}as-P\'erez, F.~Mayet, J.~Aumont}
\affil{LPSC, Universit\'e Joseph Fourier Grenoble 1, CNRS/IN2P3, Institut National Polytechnique de Grenoble, 53, av. des Martyrs, 38026 Grenoble, France}

\author{ F.-X.~D\'esert}
\affil{Laboratoire d'Astrophysique, Obs. de Grenoble, BP 53, 38041 Grenoble Cedex 9, France}



\begin{abstract}
 Whithin the framework of the Planck satellite polarisation calibration, we present a study of the Crab Nebula spectral
 energy distribution (SED) over more than 6 decades in frequency ranging from 1 to $\rm 10^6 \ GHz$
(from 299 to $\rm 2.99 \times 10^{-4} mm$). The Planck satellite
 mission observes the sky from 30 to 857 GHz (from 9.99 to 0.3498 mm) and therefore we focus on the millimetre region.
  We use radio and submillimetre data from the WMAP satellite between 23 and 94~GHz (from 13 to 3.18 mm) and from the Archeops
  balloon experiment between 143 (2.1~mm) and 545~GHz (0.55~mm), and a compendium of other Crab Nebula observations.  
  The Crab SED is compared to models including three main components : synchrotron which is  
  responsible for the emission at low and at high frequencies, dust which explains
  the excess of flux observed by the IRAS satellite and an extra component on the millimetre regime.    
  From this analysis we conclude that the unpolarised emission of the Crab Nebula at microwave and millimetre
wavelengths is the same synchrotron emission that the one observed in the radio domain. 
Therefore, we expect the millimetre emission of the Crab nebula to be polarised with the same degree of polarisation 
and orientation than the radio emission. We set upper limits on the possible errors induced by any millimetre extra component on the
reconstruction of the degree and angle of polarisation at the percent level as a maximum.
This result strongly supports the choice by the Planck collaboration of the Crab nebula emission for performing polarisation cross-checks in the range 30 (299 mm) to 353  GHz (0.849 mm). 
\end{abstract}


\keywords{Cosmic microwave background -- Cosmology: observations, calibration}



\section{Introduction}
As the strongest source of synchrotron radiation in our Galaxy, the  pulsar-powered Crab nebula ({\it Taurus A}) is a well
studied astrophysical object and it is therefore used for calibration purpose. This will be the case for the
Planck satellite mission which will use the Crab Nebula for polarisation cross-checks in the frequency
range from $30$ to $\rm 353 \ GHz$. A good understanding of the SED of the source as well as of the total intensity
flux within the Planck beam will be required for an accurate determination of the angle of polarisation of the detectors
and of a possible cross polarisation effect between detector as they  limit the accuracy to which the
CMB polarised angular power spectra will be measured. \\

The emission spectrum of the A.D. 1054 supernova 
remnant has been the subject of a host of investigations over several decades
in frequency. The radio spectrum is known to exhibit a synchrotron power law
with a spectral index $\beta \simeq -0.299 \pm 0.009$  \cite{Baars1977}.  This
 continuum from radio synchrotron seems to be fading with a rate 
$\alpha=(-0.167\pm 0.015)\% \mathrm{yr}^{-1}$ \cite{Aller1985}. At higher frequency,
above 10$^4$~GHz , the observation are also consistent with synchrotron emission with a power-law of spectral 
index $-0.73$ \cite{Veron1993}.\\
The data from IRAS satellite \cite{Marsden1984} have been reanalyzed by \cite{Strom1992} revealing a 
significant excess of emission over the low frequency synchrotron spectrum, well
explained by a single dust component at a temperature 
$\rm T \sim 46 \ K$, thus requiring a $\rm 0.02 \ M_\odot$  dust mass. 
Using MPifrR bolometer arrays at the IRAM 30 m telescope, \cite{Bandiera2002} gave the first evidence for a new component  
at milllimetre wavelengths.  They have shown that this $\rm 1.3 \ mm$  (230 GHz) excess flux cannot be interpreted as 
emission from a dust component whereas the data may be consistent with a low energy cutoff in the energy 
distribution of the emitting particles. However, at 847$\mu$m (353 GHz) \cite{Green2004} found a good agreement
with the canonical radio synchrotron emission and hence no need for an extra component. As the Planck satellite
mission will use the Crab nebula emission to perform polarisation cross-check in the range 30 to 353~GHz, a deep
knowledge of the physical origin of this emission is needed.  \\

For this purpose, we use observations of the Crab Nebula by the WMAP satellite 
at 23, 33, 41, 61 and 94~GHz \cite{Page2006} and by the Archeops balloon experiment at 143, 217, 345 and 545~GHz \cite{Macias-Perez2007,Desert2007}. 
These data in addition to already published data are used to study the
Crab Nebula SED.
The paper is organized as follows.In  Sect.~\ref{sec:archWMAP} we present the re-analysis of the
Archeops and WMAP data. Sect.~\ref{sec:EMspectrum} presents the SED of the Crab Nebula from 1 to 10$^6$~GHz
and compares it to a model including the synchrotron and dust well known components.
In Sect.~\ref{sec:modelEMspectrum} we perform a coherent analysis of the Crab SED over the full frequency range
adding an extra component to the previous model to account for the possible millimetre excess. The implications of
our results for the Planck polarisation calibration are discussed in Sect.~\ref{sec:planckpolarcalib}. Summary and
conclusions are given in Sect.~\ref{sec:conclusion}.

\section{Re-analysis of the Archeops and WMAP observations}
\label{sec:archWMAP}

\begin{figure}[h]
\begin{center}
\includegraphics[scale=0.4]{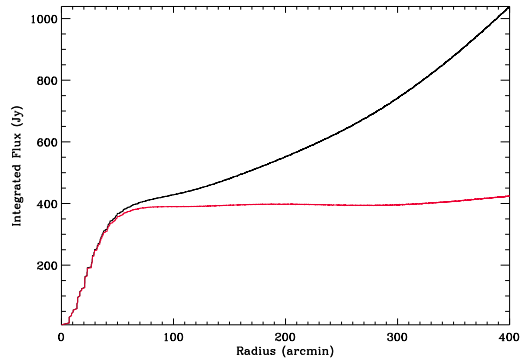}
\caption{Integrated flux at 23~GHz as a function of the angular distance to
the Crab nebula center before (black) and after (red) background
subtraction. \label{fig:backgroundradial}}
\end{center}
\end{figure}

In this paper we are mainly interested in the microwave and millimetre regime for which the best
currently available data are the WMAP \cite{Page2006} and Archeops \cite{Desert2007} data. To avoid
any bias on our analysis by a difference of treatment of these two data sets we have decided to re-estimate
the Crab nebula unpolarised emission from 23 to 545~GHz using the publicly available WMAP \cite{Page2006,Hinshaw2009}
and Archeops \cite{Macias-Perez2007} intensity maps. Furthermore, to make the full data set as homogeneous
as possible, the total intensity flux for each frequency was estimated using standard radial photometry techniques
to mimic the flux integration performed for the high resolution data sets. Notice that
we integrate far from the radio extension of the Crab nebula, $\sim$ 5 arcminutes radius, to 
recover the Crab nebula flux diluted over the entire beam pattern, Therefore, a key issue for this analysis is
the subtraction of the background galactic emission which is mainly due to synchrotron and free--free emissions at low
frequencies and dust at high frequencies. As the WMAP and Archeops maps
are available on Healpix format we have develop our own radial photometry software for this pixelization scheme. \\

The results of this analysis are presented on table~\ref{table1}. For Archeops we obtain similar values 
than those in \cite{Desert2007} with slightly different error bars. However, for WMAP we obtain values 
8 \% larger than those in \cite{Page2006}. Most probably the difference comes from the way the 
background subtraction is performed and the fact the WMAP team computed the intensity and polarisation 
simultaneously. It is important to stress that the difference with respect to the WMAP team is within the 1--$\sigma$ 
error bars at high frequency and goes up to 3--$\sigma$ at 23~GHz.
To illustrate the problem of background subraction we show on figure~\ref{fig:backgroundradial}
the radial profile centered at the Crab nebula position before (black) and after (red) background subtraction
at 23~GHz. 
For this paper we use our own estimation of the Crab nebula flux in intensity but we have noticed that
the final conclusions are not changed significantly by the choice of data set.

\section{SED of the Crab Nebula}
\label{sec:EMspectrum}
We present in this section a coherent analysis of the Crab SED in the
range 1 to 10$^6$~GHz based on a compendium of observations shown in 
Table~\ref{table1}. Notice that to be able to directly compare to the Archeops
and WMAP data, we have chosen only those data sets for which integrated
fluxes over the full extension of the Crab Nebula are available.

\begin{table*}
\begin{center}
\begin{tabular}{|c|c|c|c|c|}
\hline
$\nu$ (GHz) & $\rm S_\nu$ (Jy) & $\rm \Delta S_\nu$ (Jy) & Central Epoch &
Reference\\
\hline
\hline 
1.117 &   990.0  & 59.4   &        1969.9 &  \cite{Vinogradova1971} \\
1.304 &   980.0  & 58.8   &        1969.9 &  \cite{Vinogradova1971} \\
1.4   &   930.0  & 46.5   &        1963   &   \cite{Kellermann1969}\\
1.765 &   940.0  & 56.4   &        1969.9 &  \cite{Vinogradova1971} \\
2.0   &   840.0  & 50.4   &        1969.3 &   \cite{Dmitrenko1970} \\
2.29  &   810.0  & 48.6   &        1969.3 &   \cite{Dmitrenko1970} \\
2.74  &   795.0  & 47.7   &        1969.3 &   \cite{Dmitrenko1970} \\
3.15  &   700.0  & 24.5   &        1964.4 &   \cite{Medd1972} \\
3.38  &   718.0  & 43.1  &        1969.3 &   \cite{Dmitrenko1970} \\
3.96  &   646.0  & 38.8  &        1969.3 &   \cite{Dmitrenko1970} \\
4.08  &   687.0  & 20.6  &        1964.8 &   \cite{Penzias1965} \\
5.0   &   680  & 34   &        1963   & \cite{Kellermann1969}\\
6.66  &   577.2  & 20.2 &        1965.  &  \cite{Medd1972} \\
8.25  &   563.0  & 22.5  &        1965.9 &  \cite{Allen1967} \\
13.49 &   524.0  & 19.9 &        1969.9 &   \cite{Medd1972} \\
15.5  &   461  & 24 &        1965.9 &  \cite{Allen1967} \\
16.0  &   447.0  & 15.6 &        1970.6 &  \cite{Wrixon1972} \\
22.285&   397  & 16.0   &        1973.1 &  \cite{Janssen1974}   \\
22.5  &   395 & 7     &        2003   & {\it This paper} \\
31.41 &   387  & 72   &        1966.7 &  \cite{Hobbs1968}   \\
32.8  & 340    & 5      &        2003   & {\it This paper} \\
34.9  & 340    & 68    &        1967.3 &  \cite{Kalaghan1967}   \\
40.4  & 323  & 8.0      &        2003   & {\it This paper} \\
60.2  & 294  & 10.0      &        2003   & {\it This paper} \\
92.9  & 285  & 16.0     &        2003   & {\it This paper} \\
111,1 & 290    & 35     &        1973.5 & \cite{Zabolotnyi1976} \\
 143  & 231  & 32   &         2002  & {\it This paper}\\	  
 217  & 182 &  38  &         2002  & {\it This paper}\\
 230  & 260    & 52     &         2000  & \cite{Bandiera2002}\\
 250  & 204    &  32    &         1985.3& \cite{Mezger1986}\\
 300  & 194.0    & 19.4   &         1983  &   \cite{Chini1984} \\
 300  & 131    &  42    &         1978.75& \cite{Wright1979} \\	
 300  & 300    &  80    &         1976.0 & \cite{Werner1977}\\  
 347  & 190    &  19    &	1999.8 & \cite{Green2004} \\
 353  & 186 & 34   &         2002   & {\it This paper}\\	  
 545  & 237 & 68   &         2002   & {\it This paper}\\
 750  & 158    &  63    &         1978.75& \cite{Wright1979} \\ 
 1000 & 135    &  41    &         1978.75& \cite{Wright1979} \\	 
3000  & 184    & 13     &         1983.5 & \cite{Strom1992}\\ 
5000  & 210    & 8      &         1983.5 & \cite{Strom1992}\\ 
$\rm 12 \times 10^3$ & 67   & 4 & 1983.5 & \cite{Strom1992}\\ 	  
$\rm 25 \times 10^3$ & 37   & 1 & 1983.5 & \cite{Strom1992}\\ 	  
$3.246 \times 10^5$  & 6.57 & 0.66  & 1989 & \cite{Veron1993}\\ 
$4.651 \times 10^5$  & 4.78 & 0.48  & 1989 & \cite{Veron1993}\\ 
$5.593 \times 10^5$  & 4.23 & 0.42  & 1989 & \cite{Veron1993}\\ 
$7.878 \times 10^5$  & 3.22 & 0.32  & 1989 & \cite{Veron1993}\\ 

\hline
\end{tabular}
\caption{Compendium of Crab Nebula observations 
from  1
to $\rm 10^6 \ GHz$. Fluxes ($\rm S_\nu$) are presented in Jy. For Veron-Cetty \& Woltjer 1993 and Chini
{\it et al.} 1984 a conservative 10\%  error has been chosen to account for extrapolation errors. The central
epoch of observation is also indicated. This is used for the evaluation of the
fading effect of the low frequency synchrotron component, up to 100 GHz.
Data values labeled {\it This paper} for Archeops and WMAP are revaluated using the method described in Section 2. \label{table1}}
\end{center}
\end{table*}

\subsection{Low-frequency synchrotron emission}
\label{lowsynchrotronemission}
  
An accurate determination of the low-frequency synchrotron component 
is necessary to assess the synchrotron contribution at mm frequencies.
Any inter-comparison of low-frequency radio observations of 
the Crab nebula must take into account its well-known secular decrease. 
In particular \cite{Aller1985} have estimated a secular decrease in the flux density 
at a rate $\alpha=(-0.167\pm 0.015) \ \% yr^{-1}$ from observations 
at $\rm 8 \ GHz$ over the period 1968 to 1984. This result is in good agreement with other
studies at lower frequencies: for example $\alpha=(-0.18\pm 0.1) \ \% yr^{-1}$ over the period 1977 to 2000 at $\rm 927 \ MHz$ by 
\cite{Vinyajkin2005}. 
\begin{figure}[h]
\begin{center}
\includegraphics[scale=0.4]{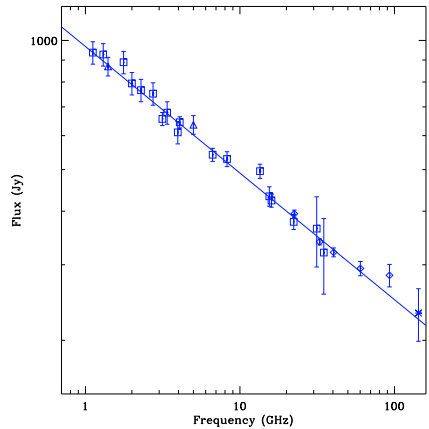}
\caption{SED of the Crab Nebula from 1 to 100~GHz.  Data samples are represented together.
with he best-fit power law model to the data (see text for details).  \label{fig:synclowfreq}}
\end{center}
\end{figure}

\begin{figure}[h]
\begin{center}
\includegraphics[scale=0.4]{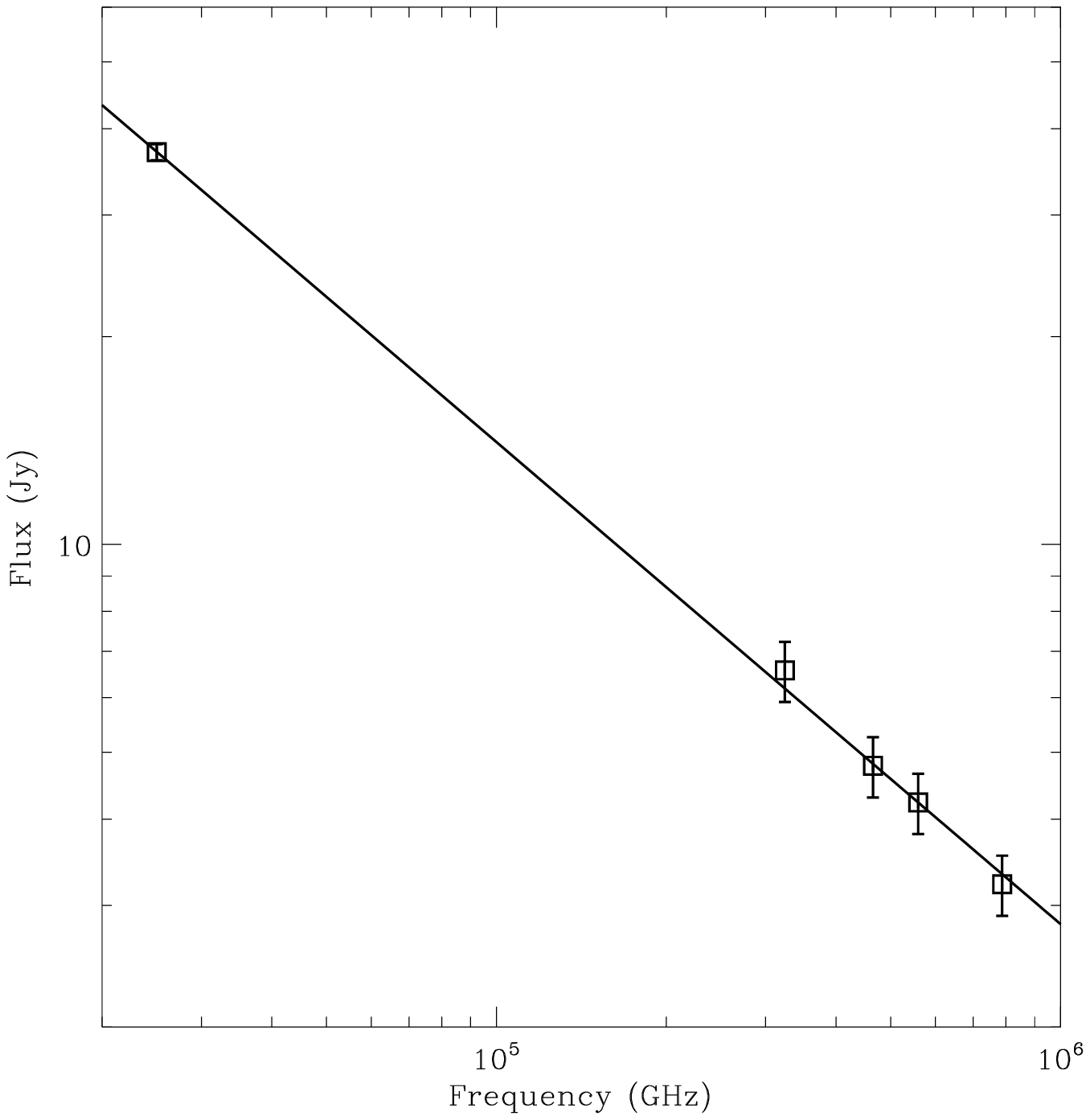}
\caption{High frequency SED of the Crab Nebula. Data samples are represented together with the.
best-fit power law model to the data (see text for details). \label{fig:synchighfreq}}
\end{center}
\end{figure}

All these measurements are in fair agreement with theoretical evaluations 
of the evolution of pulsar driven supernova remnants by \cite{Reynolds1984} which predicts $\rm \alpha $ ranging from $-0.16\%$ to $-0.4\% yr^{-1}$.\\
For this paper the value $\alpha=-0.167 \ \% yr^{-1}$ is chosen for the fading of the Crab Nebula and
all data are converted to a common observation date, 01/01/2003. In figure \ref{fig:synclowfreq} we trace
the flux of the Crab Nebula as a function of frequency for the fading corrected low-frequency data in
table~\ref{table1} ranging from 1 to 143~GHz.

We observe a large decrease of flux with increasing frequency which can be represented by a power law
of the form  $\rm A_1 \big( \frac{\nu}{1 \mathrm{GHz}} \big)^{\beta_1}$. By $\chi^2$ minimization, we obtain for the low-frequency data up to 100~GHz
$$\beta_1=-0.296 \pm 0.006 \;\textnormal{and}\; A_1 = (973 \pm 19) \ Jy$$ with $\rm \chi^2 / N_{dof} = 1.03$ and this model is shown as a
solid line on Fig~\ref{fig:synclowfreq}. This result is in agreement 
with the "canonical" value $\beta \simeq -0.299 \pm 0.009$ from \cite{Baars1977}
\footnote{\cite{Kovalenko1994} obtained $\beta_1=(-0.27\pm 0.04)$} at the 1-$\sigma$ level.

\subsection{High frequency synchrotron and dust emission}
\label{sect:highsynchrotron}

\begin{figure}[h]
\begin{center}
\includegraphics[scale=0.4]{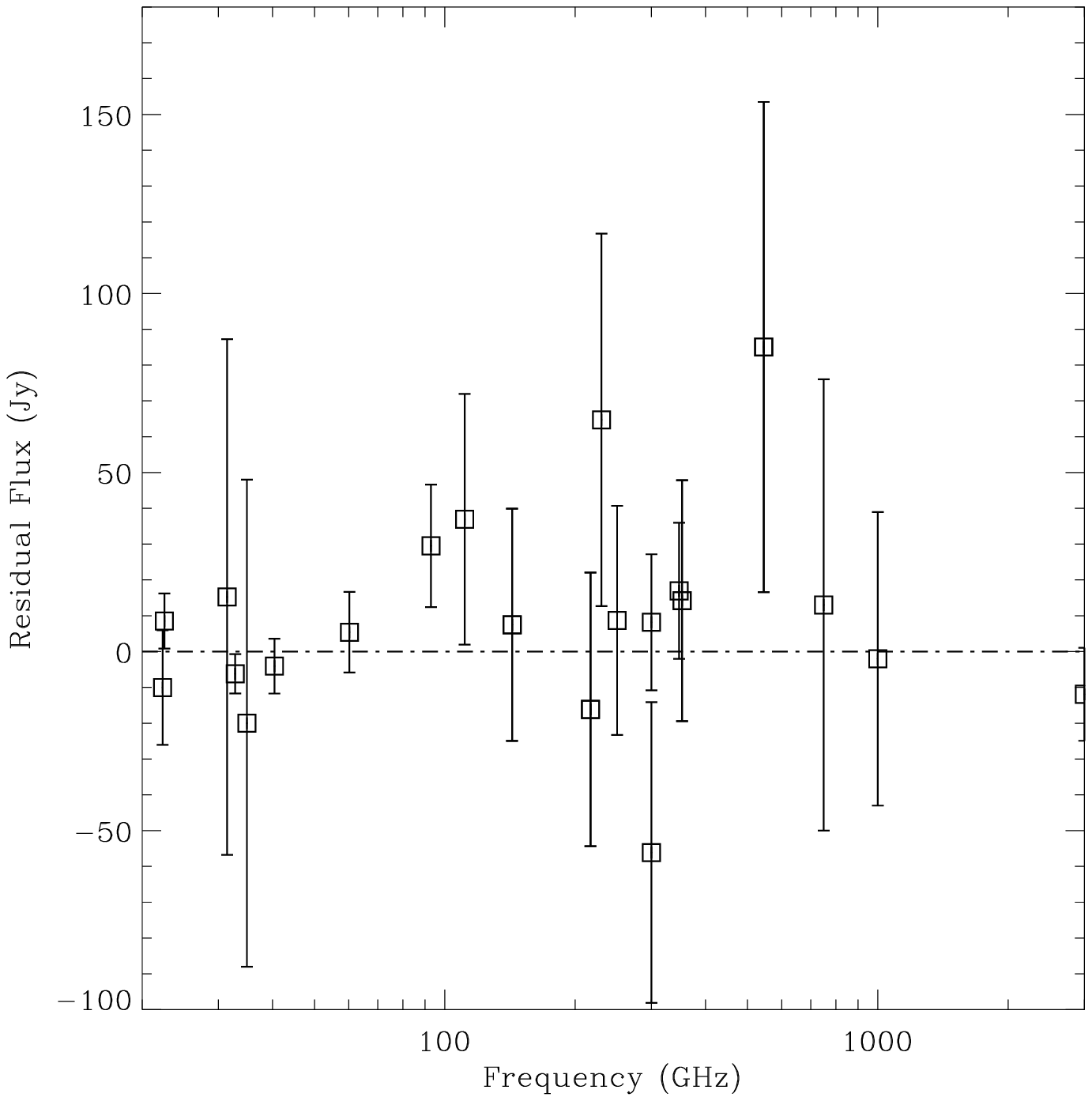}
\caption{Residual to the standard model of the Crab nebula
emission assuming a synchrotron and a dust component, in the range 10 to 2000~GHz (see text for details).
Notice that the millimetric data were not used in the fit. We show the full data set in the region of interest.
\label{fig:mmexcess}}
\end{center}
\end{figure}
The Crab synchrotron emission described above evolves at higher frequencies, around $10^4$~GHz, towards a
much harder SED with a spectral index of $\sim -0.73$ (see for example \cite{Veron1993}).
To accurately estimate the synchrotron emission properties at high frequency we have fitted the data
from $10^4$ to $\rm 10^6 \ GHz$ presented in table~\ref{table1} to a power law of the form $\rm A_2 \nu^{\beta_2}$.
By $\chi^2$ minimization, we found the data to be well fitted, $\rm \chi^2 / N_{dof} = 0.155$, by a power law of parameters
$$\rm \beta_2=(-0.698\pm 0.018) \;\textnormal{and}\; A_2 = (43.5  \pm 8.8)\times 10^3 \, Jy$$
Figure~\ref{fig:synchighfreq} represents the high-frequency data and the best-fit power law model is overplotted. \\

Finally, the infrared satellite observatory IRAS has revealed significant excess emission above this synchrotron 
contribution around 50 $\rm \mu m$ \cite{Marsden1984}. As shown by  \cite{Strom1992} this can be explained, 
after careful removing of the synchrotron component, by a single dust component described by a modified
black body of emissivity $\beta = 2$ at T$\rm  = 46 \pm 3 \ K$, requiring a dust mass of $\rm 0.02 \ M_\odot$.  

\subsection{Millimetric excess}
\label{sect:mmexcess}

To evaluate a possible millimetric excess of flux in the range 100 to 1000~GHz we assume the above canonical modeling 
of the Crab nebula SED: a synchrotron component with a spectral index break at high frequency and a dust component
at infrared wavelengths. We can thus reconstruct the Crab nebula emission at the millimetric
frequencies and compare it to the data in table~\ref{table1}.
Fig~\ref{fig:mmexcess} shows the residuals to the canonical model from 10 to $2 \times 10^4$~GHz. We observe that there is no significant excess of power in the millimetric regime from 100 to 1000~GHz except for the 545~GHz 
Archeops data sample which presents only a 1.5 sigma excess. Indeed, the $\chi^2/ N_{dof}$ for the null hypothesis is $0.69$.

\section{Refined modeling}
\label{sec:modelEMspectrum}
In the previous section we have proved that the millimetric data in table~\ref{table1},
from 100 to 1000~GHz, are compatible with the canonical model assuming single synchrotron and
dust components. However, it is interesting to check if an extra component may improve significantly the
fit to the data. Following \cite{Bandiera2002} we consider either an extra low-temperature
dust emission or an extra synchrotron component. Thus, the three component model is defined as
follows
\begin{enumerate}
\item Canonical synchrotron that is described by four parameters: spectral index and amplitude for
the low and high frequency emission. At high frequency we consider both the amplitude and the
spectral index fixed and set them to the values obtained in Sect.~\ref{sect:highsynchrotron}. At low
frequency we fix the spectral index to the value obtained in  Sect.~\ref{lowsynchrotronemission} and 
the amplitude, $A_{S}$, is fitted. We also assume a constant fading of $\alpha=-0.167 \ \% yr^{-1}$
as before.

\item Canonical dust (following \cite{Strom1992}) described by a modified black body with three free parameters,
$A_{D}$, $T_{D}$ and $\beta_{D}$ that represent the amplitude, temperature and spectral index respectively.

\item One of the extra components as described below.
\end{enumerate}

\subsection{Extra low-temperature dust emission }
\label{sect:extralowtempdust}

\begin{figure*}
\begin{center}
\includegraphics[scale=0.5]{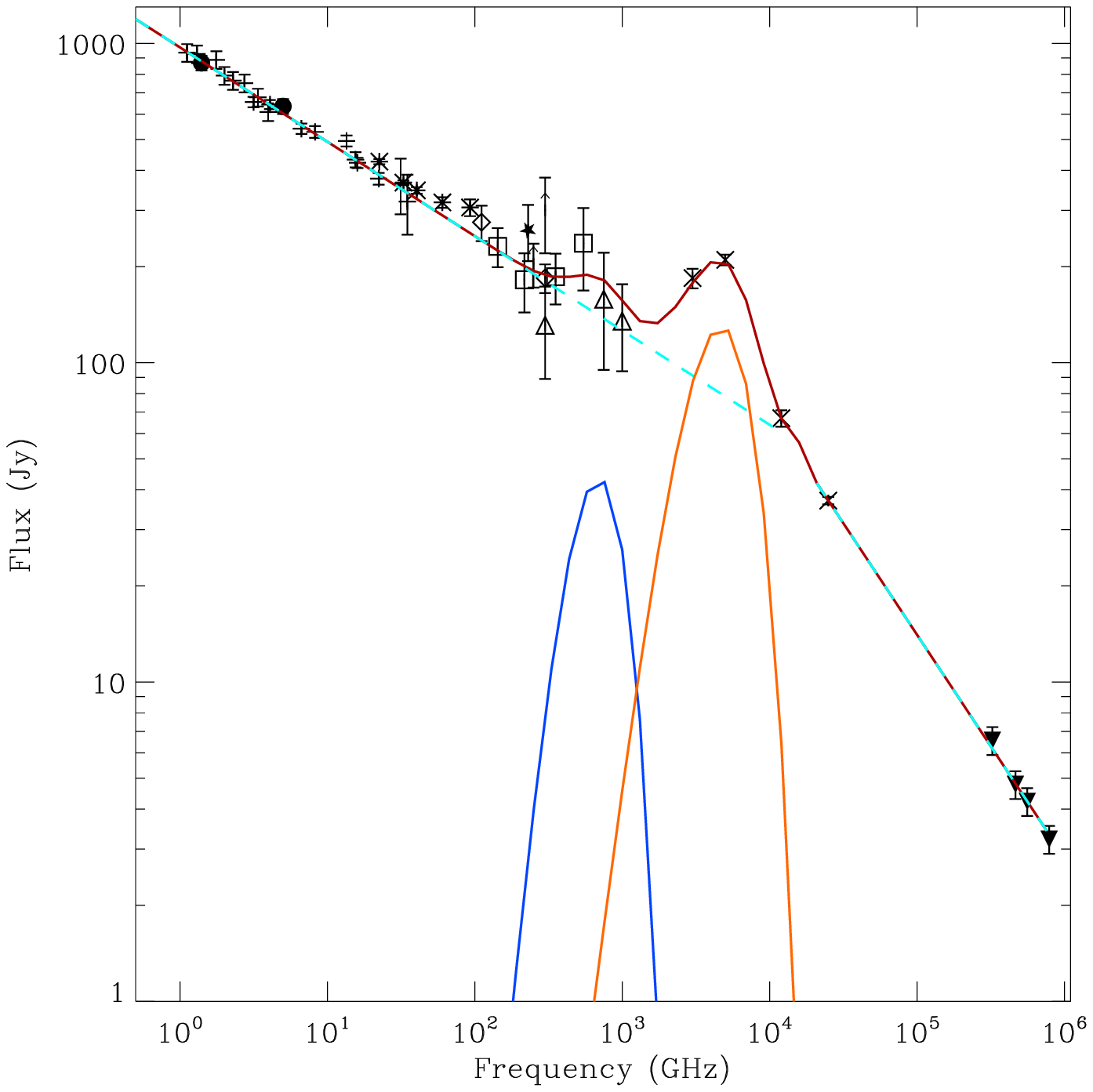}
\includegraphics[scale=0.5]{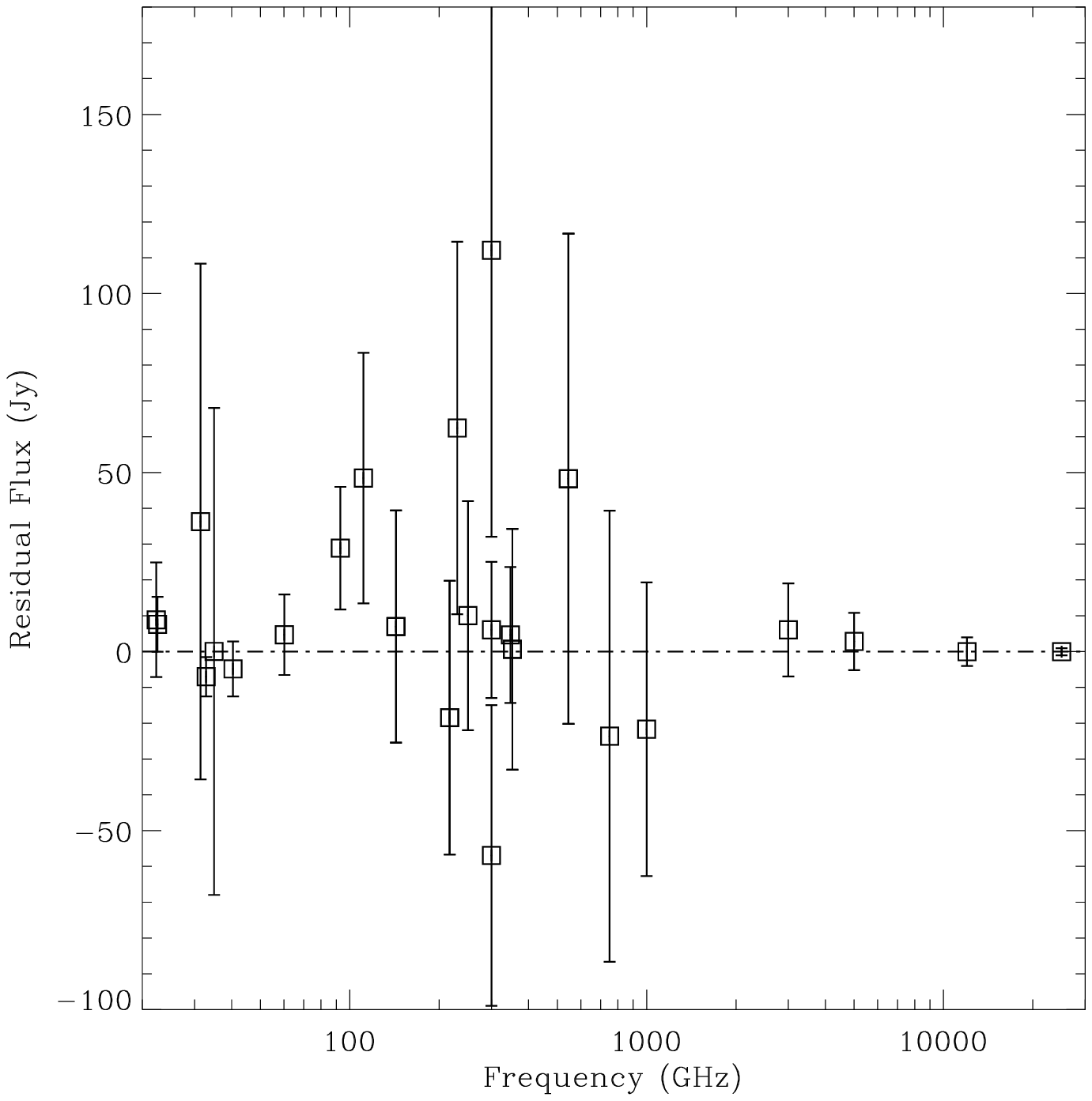}

\caption{Refined modelling of the SED of the Crab nebula in the frequency range from 1 to 10$^6$~GHz assuming an
extra dust component at low-temperature. 
Left panel: We represent the data from table~\ref{table1} (black), the best-fit model to the data (red)
and the synchrotron (light blue), the main dust (orange)  and the extra dust (dark blue) components associated to it. Right panel: residuals in the
millimetric regime for the top panel models.
\label{fig:dustfit}}
\end{center}
\end{figure*}

\begin{table*}
\begin{center}
\begin{tabular}{|c|c|c|c|c|c|c|c|c|c|}
\hline
$\chi^2 / N_{dof} $ & $A_{s} \ \mathrm{(Jy)}$ & $A_{D} \ \mathrm{(Jy)}$  & $T_{D} \ \mathrm{(K)}$  & $\beta_{D}$ & $A_{LTD} \ \mathrm{(Jy)}$  & $T_{LTD} \ \mathrm{(K)}$  &  $\beta_{LTD}$ & LTD \% & $\chi^2_{mm} / N_{dof}$  \\
\hline
0.73 & $971 \pm 9$ & $128 \pm 8$ & $ 45.9 \pm 1.2$ & $ 1.93 \pm 0.50$ &  $32 \pm 24$ & $5 \pm 23$ & $ 3.6 \pm 2.4$  & $26 \pm 20$ & 0.88 \\
\hline
\end{tabular}
\caption{Best-fit model parameters and errors for the extra dust component model. $\chi^2/N_{dof}$
 for the full data set and on the millimetric regime from 100 to 1000~GHz. LTD \% stands for the percentage of flux at 545~GHz due to the extra dust component with respect to the total flux. \label{table2}}
\end{center}
\end{table*}

\begin{figure*}
\begin{center}
\includegraphics[scale=0.5]{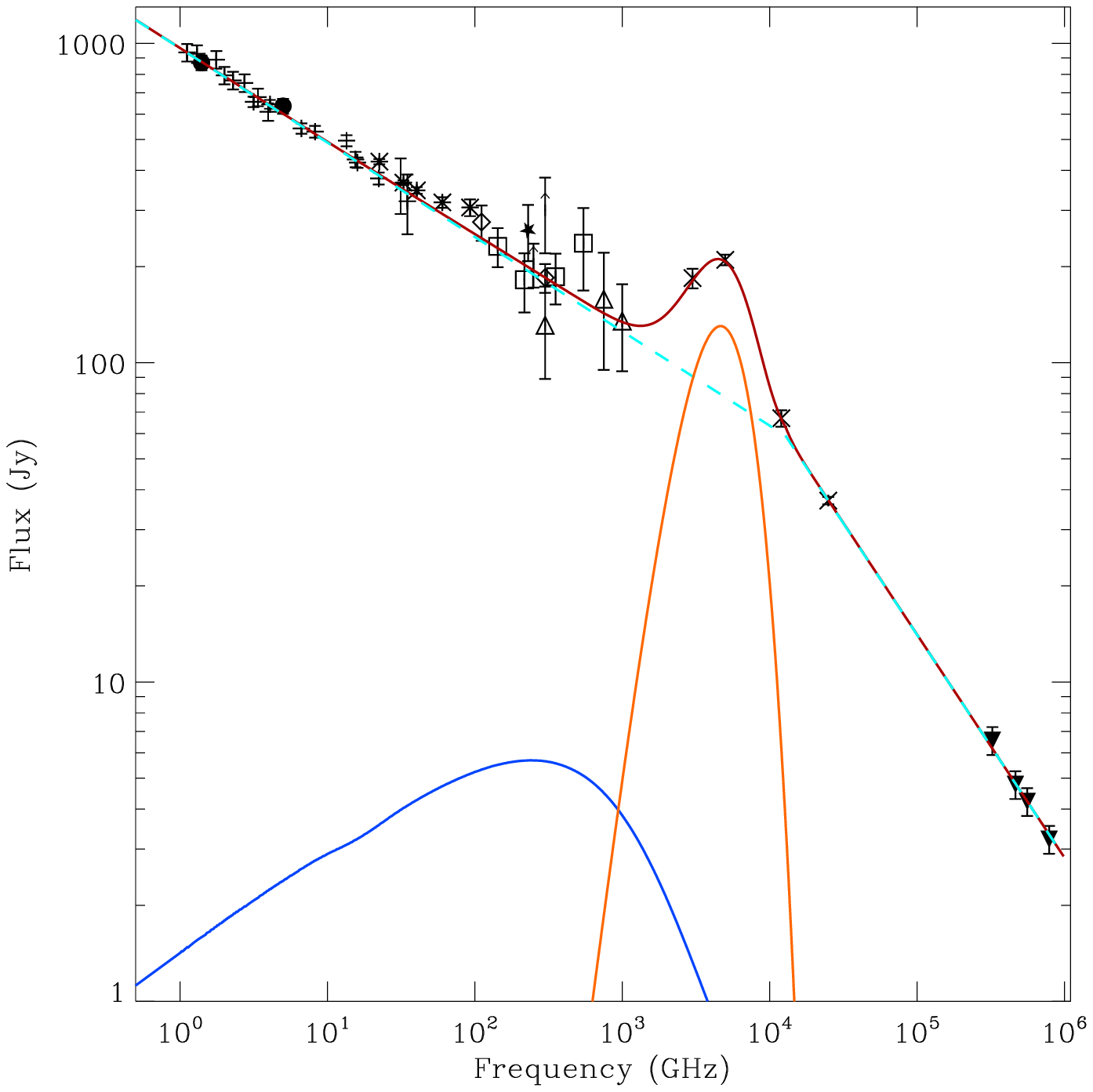}
\includegraphics[scale=0.5]{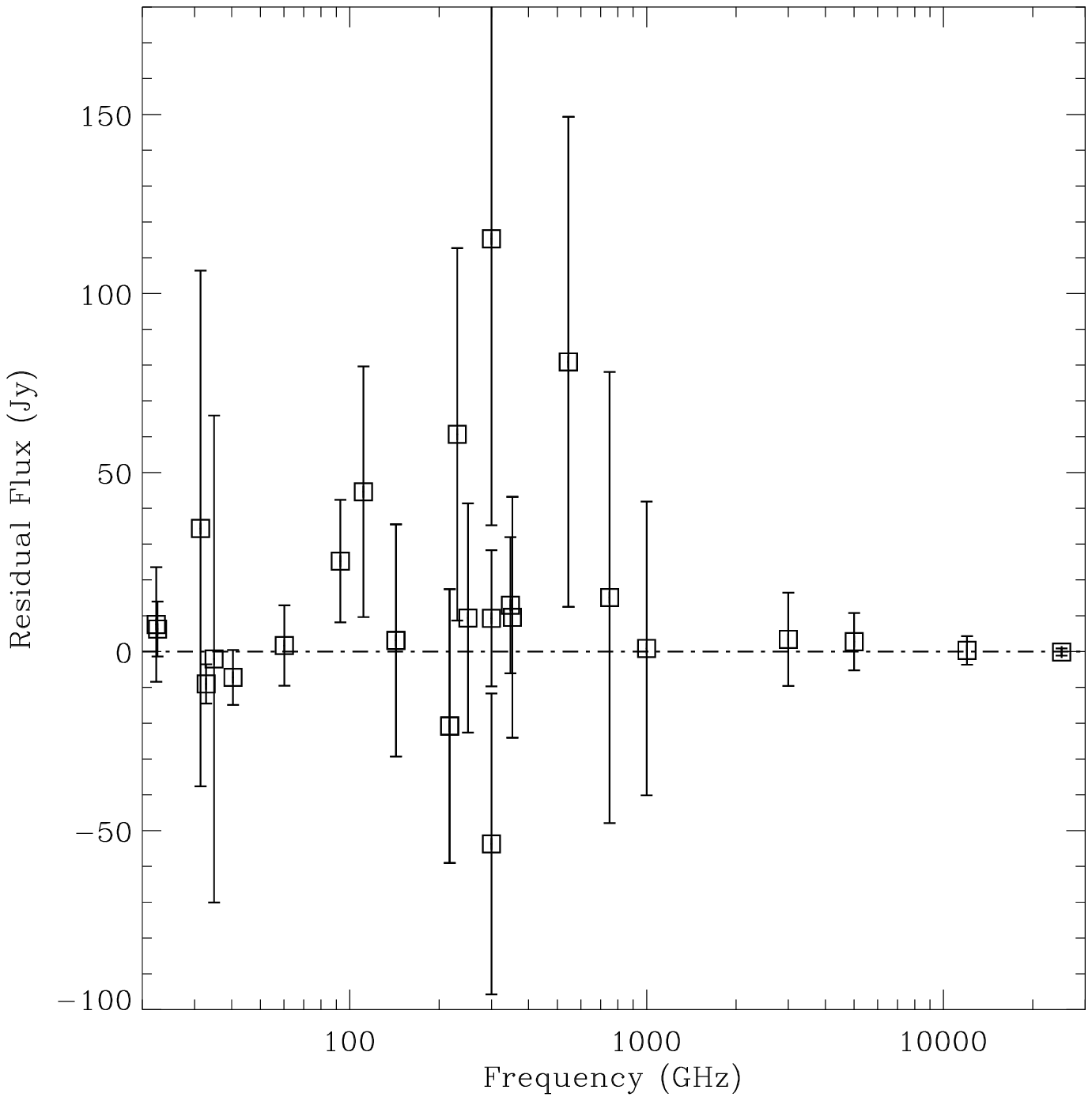}
\caption{Refined modelling of the SED of the Crab nebula in the frequency range from 1 to 10$^6$~GHz assuming an
extra synchrotron component. Left panel: We represent the data from table~\ref{table1} (black), the best-fit model to the data (red)
and the main synchrotron (light blue), the extra synchrotron (dark blue) and the dust components (orange) associated to it. Right panel: residuals in the
millimetric regime for the top panel models. 
\label{fig:lowsyncfit}}
\end{center}
\end{figure*}

\begin{table*}
\begin{center}
\begin{tabular}{|c|c|c|c|c|c|c|c|c|c|}
\hline
$\chi^2/N_{dof}$ & $A_{s} \ \mathrm{(Jy)}$ & $A_{D} \ \mathrm{(Jy)}$  & $T_{D} \ \mathrm{(K)}$  & $\beta_{D}$ & $A_{LECS} \ \mathrm{(Jy)}$  & $\nu_{c} \ \mathrm{(GHz)}$  &  $p$ & LECS \% &  $\chi^2_{mm} / N_{dof}$ \\
\hline
$0.76$ & $965 \pm 14$	& $129 \pm 9$  & $46.0 \pm 1.0$ & $1.89 \pm 0.4$ & $5 \pm 6$ & $470 \pm 317$  & $3 \pm 12$  & $ 3 \pm 3$  & 0.93 \\
\hline
\end{tabular}
\caption{Best-fit model parameters and errors for the extra syncrotron component model. $\chi^2/N_{dof}$
 for the full data set and on the millimetric regime from 100 to 1000~GHz. LECS \% stands for the percentage of flux at 545~GHz due to the extra
synchrotron component with respect to the total flux. \label{table3}}
\end{center}
\end{table*}

In Sect.~\ref{sect:mmexcess} we concluded that at $\rm 545 \ GHz$ the difference between the standard
model and the data is at its maximum. In the case of an extra dust component this implies
very low temperature dust (LTD) in the range from 4 to 10~K. To model this component we assume a modified black
body spectrum with three free parameters $A_{LTD}$, $T_{LTD}$ and $\beta_{LTD}$ that represent the 
amplitude, temperature and spectral index respectively. The best-fit model to the data is found
by $\chi^{2}$ minimization on the full frequency range from 1 to 10$^6$~GHz. \\

Figure~\ref{fig:dustfit} presents on the left panel the observational data (black)
and the global best-fit model to the data (in red). The canonical synchrotron component
is shown on light blue and the canonical dust in orange. The extra low temperature dust component
is represented in blue. The right panel presents the residuals to the best fit-model in the frequency
range of interest. The parameters and error bars for the best-fit model to the data
and the percentage of flux at 545~GHz associated to the extra component with respect to the total flux 
are given in table~\ref{table2}. We also present the $\chi^2 / N_{dof}$ 
values for the global fit in the frequency range from 1 to 10$^6$~GHz and on the millimetric 
range from 100 to 1000~GHz.\\

We obtain a good global fit to the data as shown by the $\chi^2 / N_{dof}$  value. The best-fit parameters obtained for the canonical synchrotron and dust components are in good agreement
with those presented in Sect.~\ref{sec:EMspectrum}. Comparing with  \cite{Strom1992}, we found the same dust temperature, $46 \pm 1$~K, with an  error bar 
improved by a factor of three, as we carefully account for the canonical synchrotron spectrum.
The amplitude of the extra component is compatible with zero and therefore we conclude that
there are no evidence for an extra component in the form of low--temperature dust.
Furthermore, we observe that on the one hand the data require extremely low temperatures of $5$~K with rather
large error bars, therefore dust masses of $\rm \sim 230 \ M_\odot$, making the model rather unrealistic. 
On the other hand, the improvement of the $\chi^2 / N_{dof}$ in the millimetric region between
100 and 1000~GHz is not significant to justify the addition of the three extra parameters
required by the low-temperature dust component. This is also clear on the right panel of
 figure~\ref{fig:dustfit} where we represent  the residuals to the best-fit
model to the data on the millimetric regime.

\subsection{Extra synchrotron component}

For the extra synchrotron component we consider, as in~\cite{Bandiera2002}, that the
distribution of energy of the relativistic electrons responsible for the emission 
is well represented by a power law with spectral index in the range 1 to 3 and present a low-energy
cutoff. To account for an excess of flux in the millimeter regime, the critical frequency
corresponding to the lowest energy electrons must be somewhere in the range 200 to 600~GHz.
In total the low-energy cutoff synchrotron (LECS) model has three parameters, $p$ the spectral
index of the electron energy distribution, $\nu_{c}$ the low-energy cutoff critical
frequency and $A_{LECS}$, a normalization coefficient. The best-fit to the data is
found by  $\chi^2$ minimization. \\

The parameters and error bars of the best-fit to the data for the three data sets
as well as the percentage of flux due to the extra component with respect to the total
flux at 545~GHz are given on table~\ref{table3}. We also present $\chi^2/N_{dof}$ values 
for the best-fit to the data on the full data set from 1 to 10$^6$~GHz and on the
millimetric regime from 100 to 1000~GHz.The left panel of 
figure~\ref{fig:lowsyncfit} presents the data in black
and the global best-fit model to the data is represented in red. The canonical synchrotron component
is shown on light blue and the canonical dust in orange. The  extra low-energy cutoff synchrotron component
is represented in blue. \\

As above, we obtain a good global fit to the data as shown by the $\chi^2 / N_{dof}$  value.
The best-fit parameters obtained for the canonical synchrotron and dust components are in good agreement
with those presented in Sect.~\ref{sec:EMspectrum} and on \cite{Strom1992}.
The amplitude of the extra synchrotron component is compatible with zero as well as
the value of the spectral index of the electron energy distribution. We therefore conclude that
there is  no evidence of an extra component in the form of low-energy cutoff synchrotron.
In addition, we observe that the $\chi^2/ N_{dof}$ in the millimetric region from
100 to 1000~GHz is worse than the one obtained in Sect~\ref{sect:mmexcess} assuming the canonical model only.
The right panel of figure~\ref{fig:lowsyncfit} represents the residuals to the best-fit
model to the data on the millimetric regime. From this we conclude
that the fit to the data in this frequency regime is not improved by adding an extra synchrotron component. \\

We have also performed the analysis of the data set considering the $p$ and $\nu_{c}$ parameters fixed and set to the values
quoted by \cite{Bandiera2002}. The analysis shows that the amplitude of the extra synchrotron component is compatible with zero  and therefore we conclude that the data show no evidence of an extra synchrotron component.

\section{Implications for the Planck satellite polarisation calibration}
\label{sec:planckpolarcalib}

\begin{figure*}
\begin{center}
\includegraphics[scale=0.35]{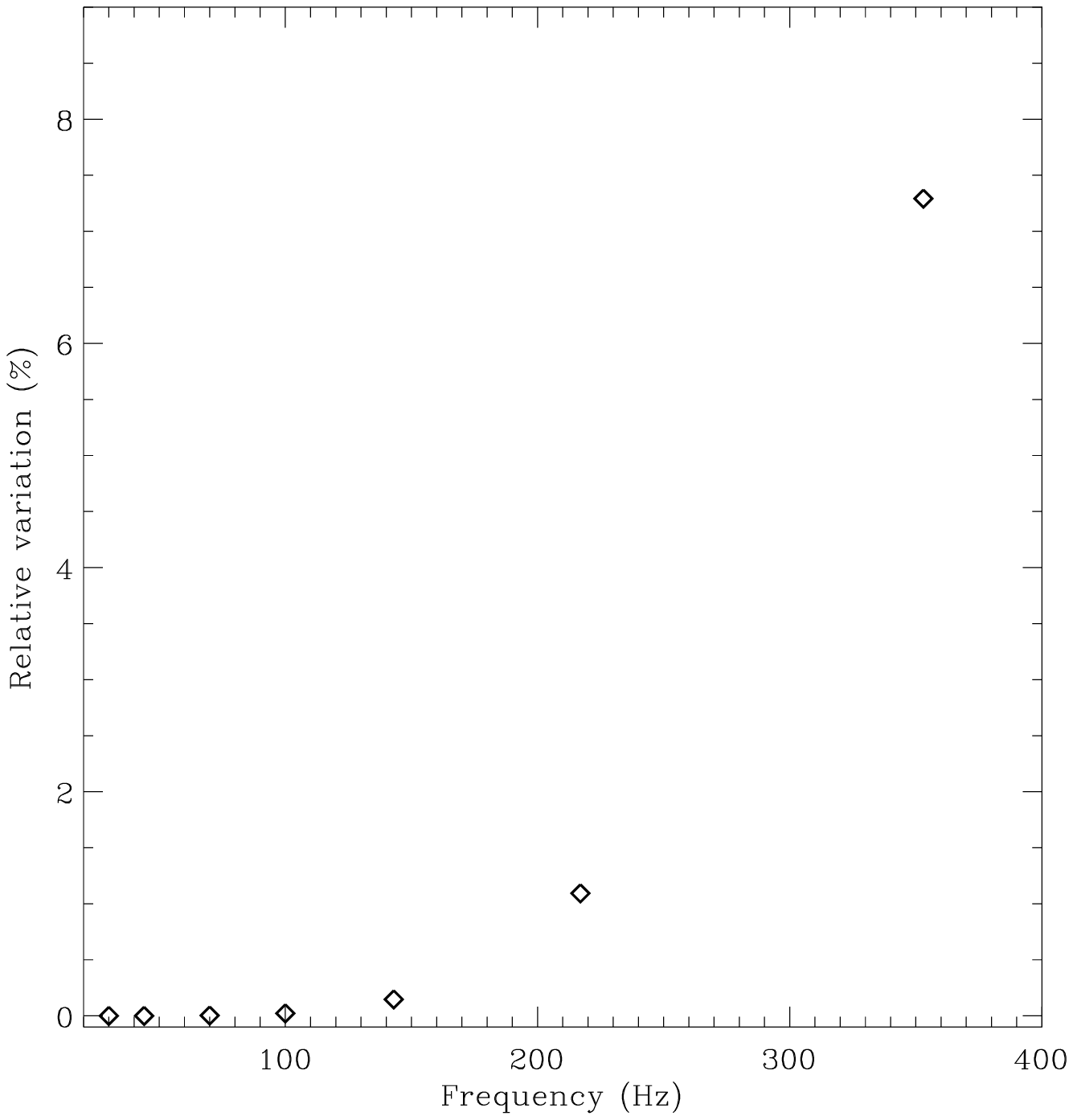}
\includegraphics[scale=0.35]{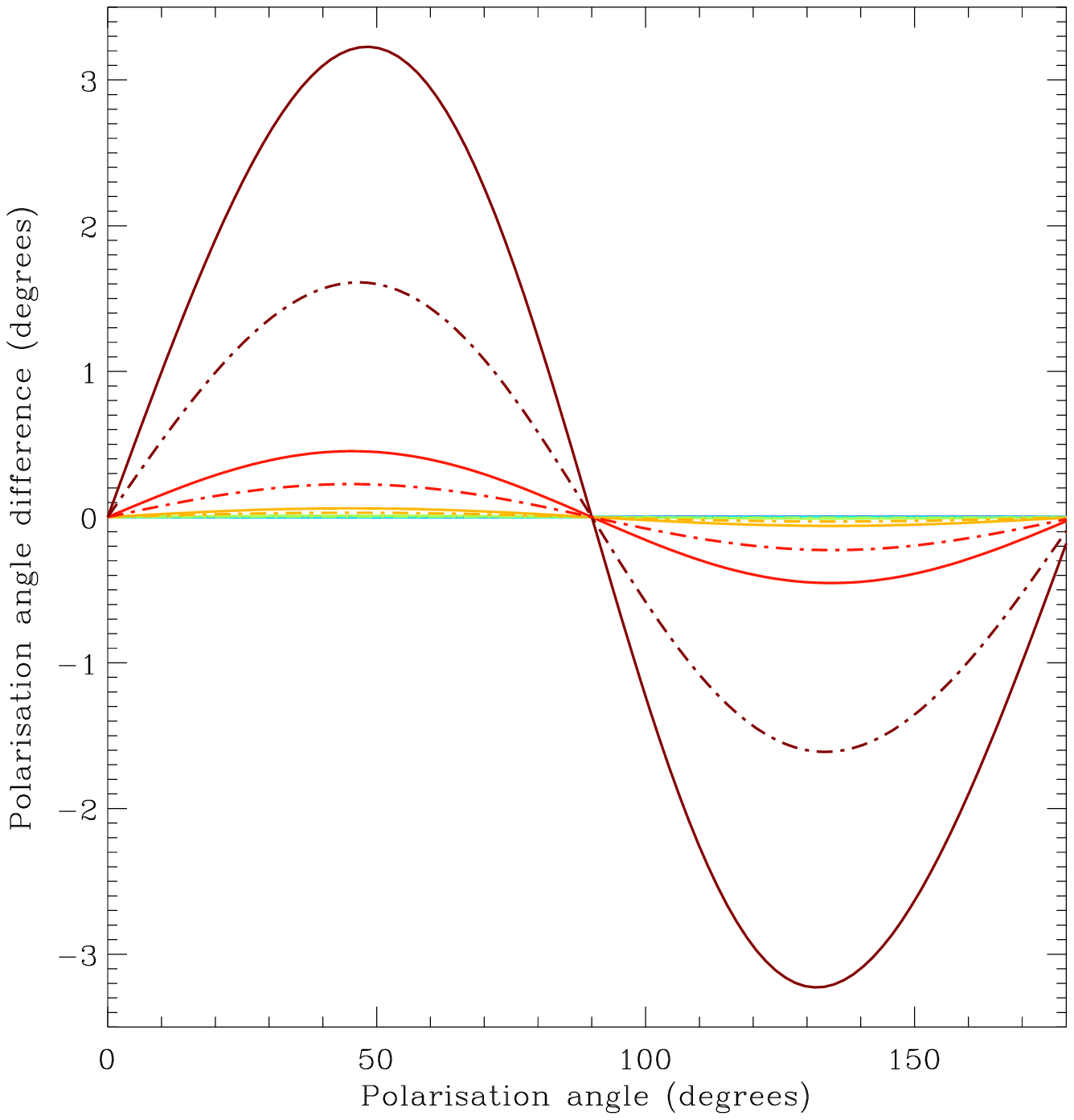}
\includegraphics[scale=0.35]{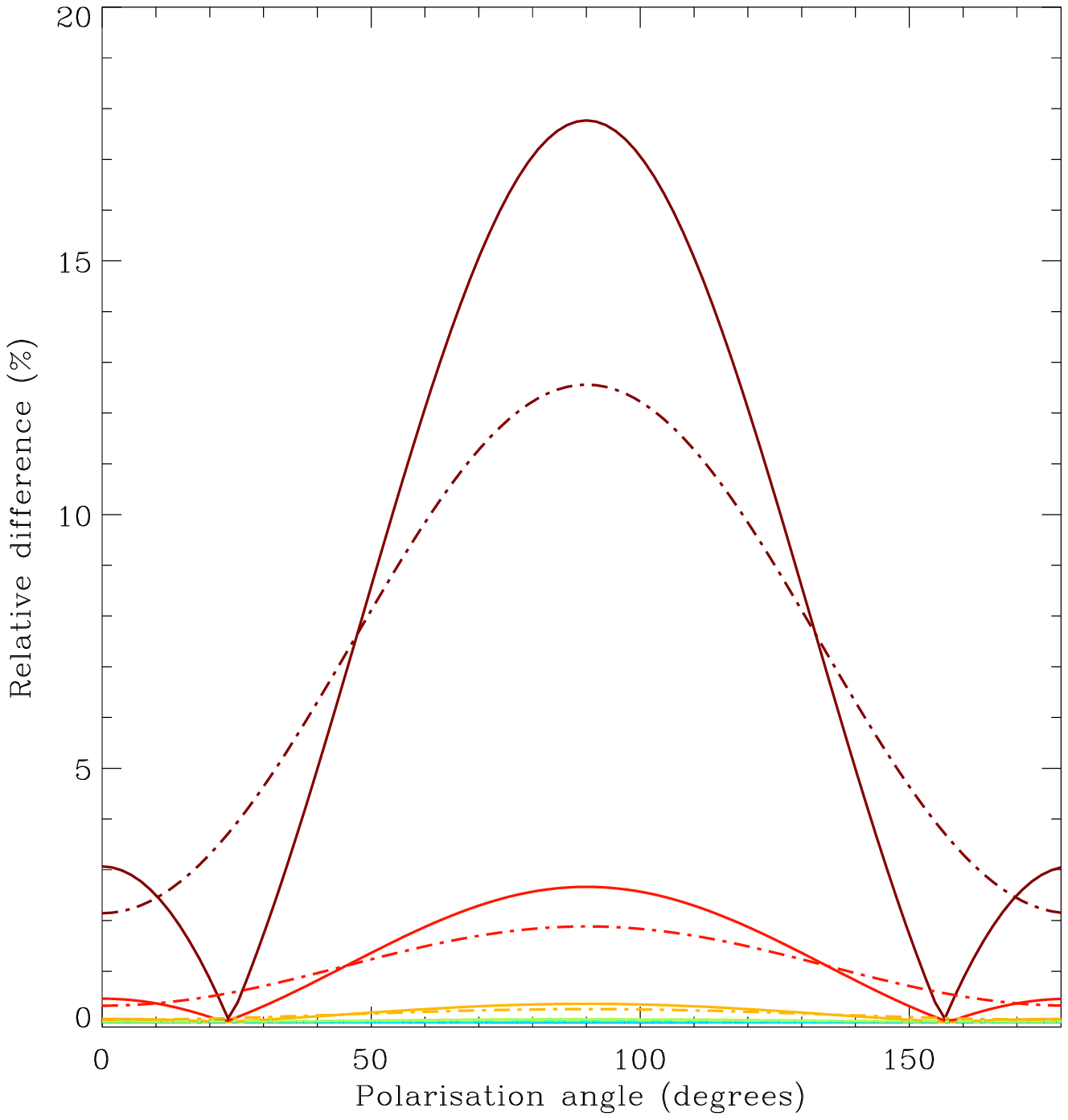}
\caption{Errors induced in the degree and angle of polarisation by an extra low--temperature
dust component. Left plot: relative error on the degree of polarisation on the case of an 
unpolarised extra component as a function of the frequency of the polarised Planck channels. Middle plot and
right plot: relative error on the degree of polarisation  and polarisation angle error on the case
of an extra polarised component as a function of the polarisation angle of this component. The dashed and solid lines
correspond to an extra component with a degree of polarisation of 5 \% and 10 \% respectively. From blue to red we 
represent the values for the different polarised Planck channels from 30 to 353~GHz.
\label{fig:dustpolerr}}
\end{center}
\end{figure*}

\begin{figure*}
\begin{center}
\includegraphics[scale=0.35]{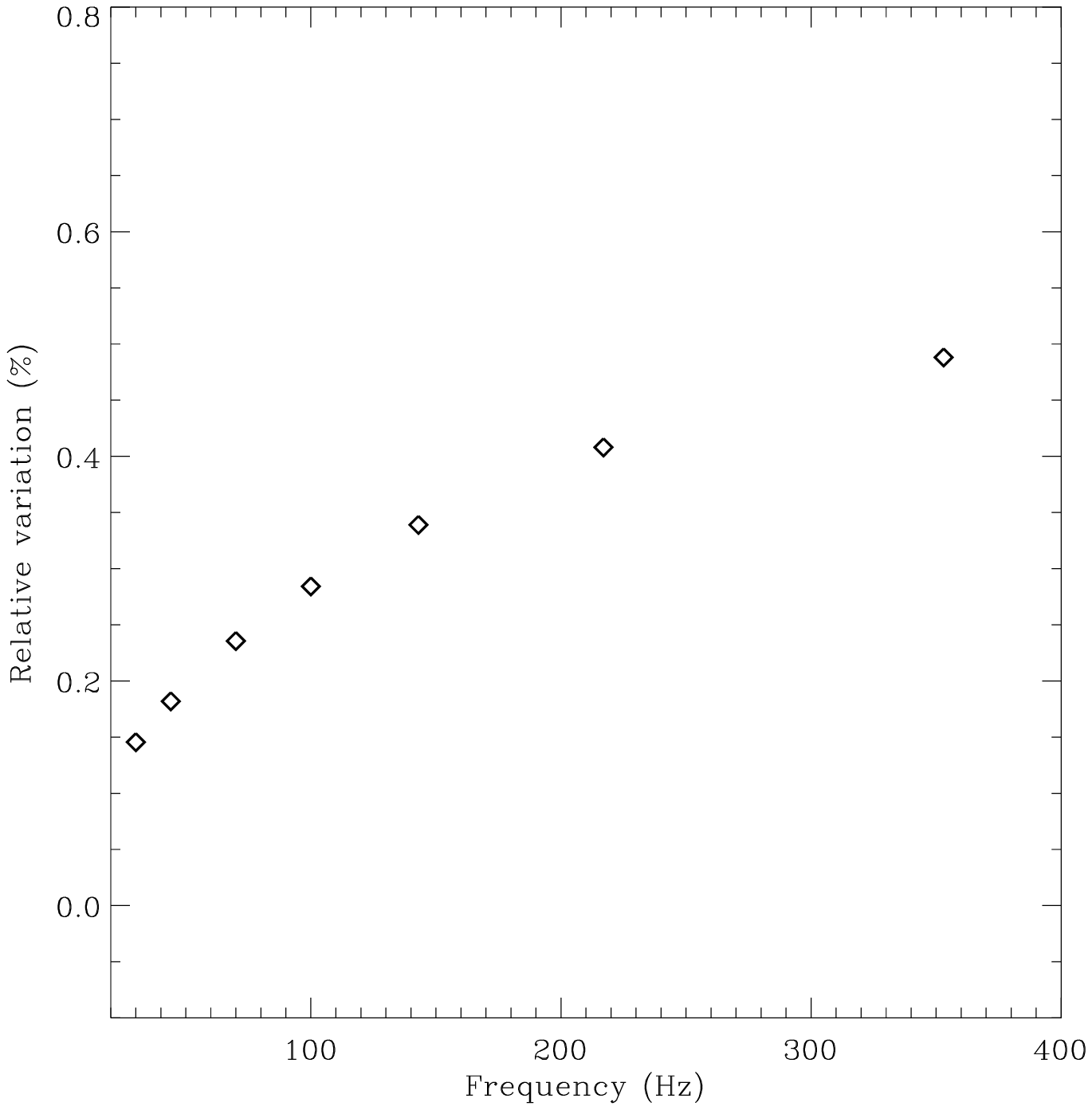}
\includegraphics[scale=0.35]{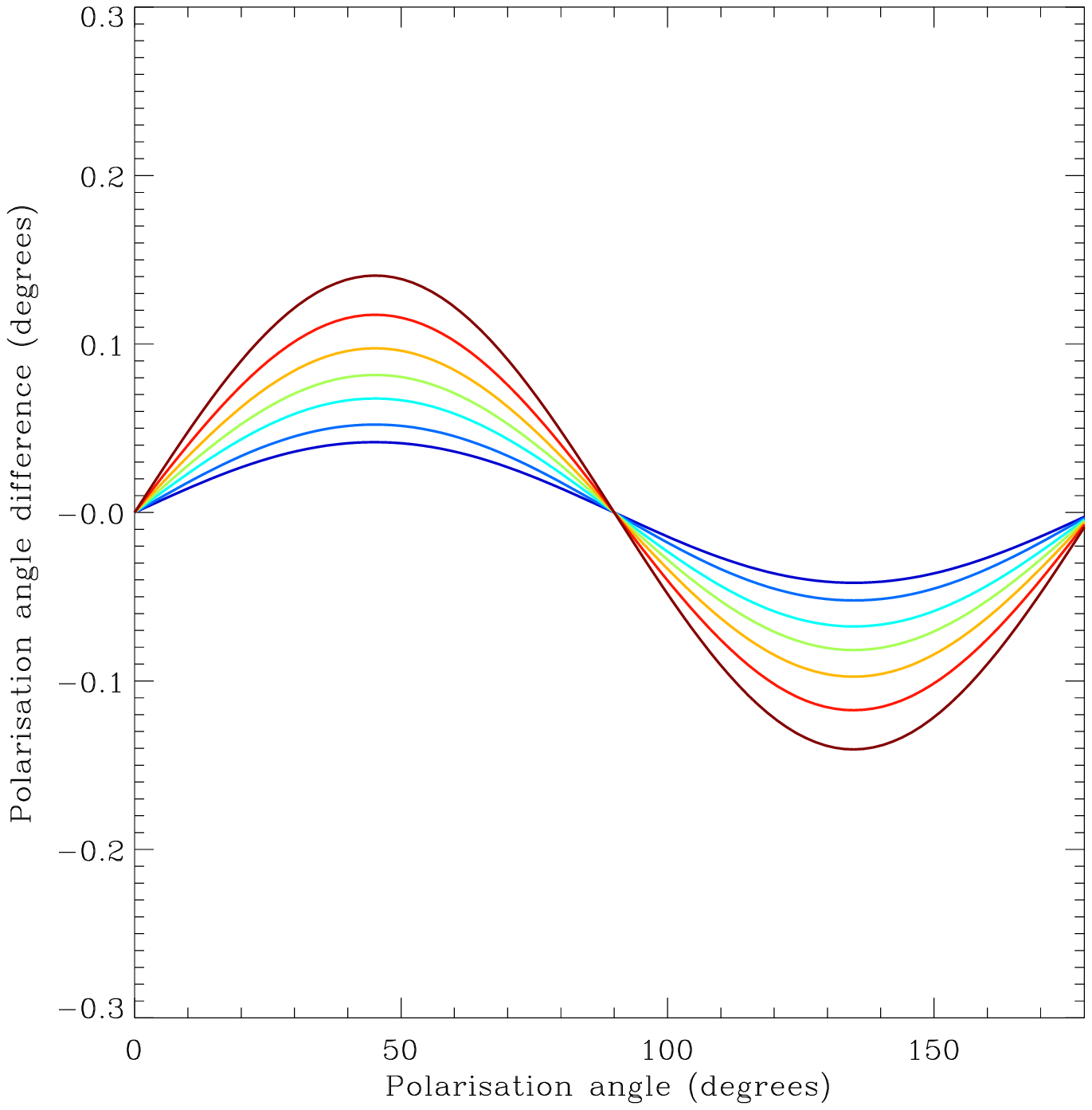}
\includegraphics[scale=0.35]{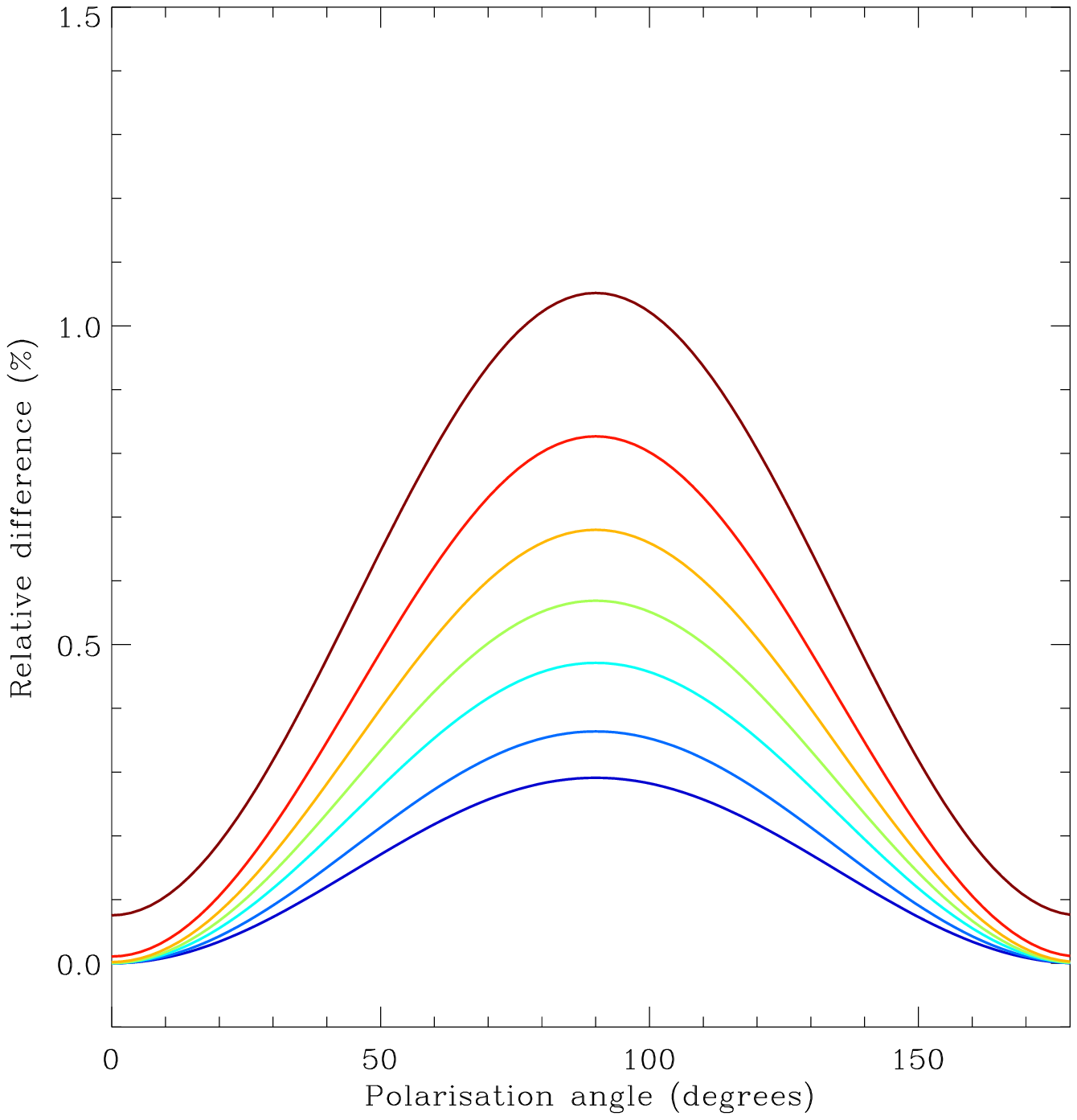}
\caption{Errors induced in the degree and angle of polarisation by an extra low--energy cutoff synchrotron
component. Left plot: relative error on the degree of polarisation on the case of an 
unpolarised extra component as a function of the frequency of the polarised Planck channels. Middle plot and
right plot: relative error on the degree of polarisation  and polarisation angle error on the case
of an extra polarised component as a function of the polarisation angle of this component. From blue to red we 
represent the values for the different polarised Planck channels from 30 to 353~GHz. 
\label{fig:lowsyncfixedpolerr}}

\end{center}
\end{figure*}

From the previous section, we conclude that the intensity emission of the Crab nebula at millimetre and submillimetre 
wavelengths is dominated by the well known synchrotron radiation observed at radio wavelengths. The data show no evidence
of an extra synchrotron component and evidence for an extra dust component is not
significant. Therefore, we can postulate that the Crab nebula emission at millimetre and 
submillimetre wavelengths is produced by the same relativistic electrons that the ones producing the radio emission. Then 
it is reasonable to guess that Crab nebula presents the same polarisation properties at radio and millimetre wavelengths. 
These conclusions are also supported 
by the Crab polarised emission observed from 23 to 94~GHz by the WMAP satellite \cite{Page2006} when compared 
to the 363~GHz SCUBA measurements \cite{2003MNRAS.340..353G} and the \cite{1991MNRAS.249.4P} 273~GHz data.
From these, it makes sense to use the low--frequency observations of the Crab nebula
to cross-check and eventually update the  knowledge of the polarisation characteristics of 
the high--frequency instruments as proposed by the Planck collaboration. As a summary and using the results from \cite{Page2006,2003MNRAS.340..353G,1991MNRAS.249.4P} we expect the Crab nebula emission at the Planck §frequencies to be polarised with a degree
of polarisation of 8 to 9 \% and a polarisation angle of 150$^\circ$ in equatorial coordinates \cite{Aumont2009}. 
Then, at 30~GHz and 353~GHz
we expect the total intensity to be $351 \pm 7$ and $173 \pm 5$ Jy and the polarised intensity about 
28 and 15 Jy, respectively. \\

However, we can estimate upper limits on the errors induced on the determination of the degree of polarisation
and polarisation angle of the Crab nebula emission at the Planck observation frequencies by the presence of an extra component.
For this we use the results obtained on the previous section and we assume two different cases: polarised and unpolarised
extra component. Figure~\ref{fig:dustpolerr} shows 1-$\sigma$ upper limits on the case of a low--temperature dust
extra component. For the left plot we assume an unpolarised extra component and we present the error on the determination
of the degree of polarisation as a function of the observation frequency. This error is below 1 \% for the Planck CMB
channels from 70 to 217~GHz and rise up to 8\% at 353~GHz. The middle and right plot assume a polarised extra component
and show the error on the polarisation angle and degree of polarisation, respectively, as a function of the polarisation angle of
the extra component with respect to the main synchrotron component. From blue to red we show the errors from 30 to 353~GHz.
The dotted-dashed and solid curve assume a degree of polarisation of 5 and 10 \% for the extra component. As before the error
on the degree of polarisation is below 1 \% at the CMB channels and from 12 to 18 \% at 353~GHz. For the polarisation
angle the error is below 0.2$^\circ$ for the CMB channels and at maximum of 3$^\circ$ at 353~GHz. Notice that the increase
in the errors at 353~GHz is mainly due to the large value and large error bars of the Archeops data at 545~GHz and 
we do not think it is significant. Figure~\ref{fig:lowsyncfixedpolerr} shows the errors induced by an extra low-energy
cutoff synchrotron component
either unpolarised or polarised with a degree of polarisation equivalent to the one expected for the canonical synchrotron
component. In this case the errors are much lower being well below 1 \%  and 0.2$^\circ$ for the degree and angle of polarisation,
respectively, at all polarised Planck frequencies.

\section{Summary and conclusions} 
\label{sec:conclusion}
Within the framework of the polarisation calibration of the Planck satellite emission
we present in this paper a global analysis of the SED of the Crab
Nebula in the frequency range from 1 to 10$^6$~GHz. For this purpose we
have used new data from the WMAP satellite \cite{Page2006} and from the Archeops balloon experiment~\cite{Desert2007}
in addition to data currently available. We focus on the centimetric and  millimetric regime as Planck observes the sky
from 30 to 857~GHz. \\



We have first shown that the data are compatible with the canonical model of Crab nebula emission which 
assumes a synchrotron component at low and high frequencies, {\it plus}
a dust component at the micrometre wavelengths.
To check if an extra component may improve significantly the
fit to the data, we consider either an extra low-temperature
dust emission or an extra low-energy cutoff synchrotron component, as in \cite{Bandiera2002}.
In both cases, we conclude that the current data 
present no evidence of an extra component. Therefore, the un-polarised millimetre flux of
the Crab nebula is well represented by the following synchrotron power--law model
\begin{equation}
 F_{\nu} = 973 \ \pm \ 19  \, \mathrm{Jy} \  {\big(\frac{\nu}{\mathrm{1 \ GHz}}\big)}^{(-0.296 \ \pm \ 0.006)}  \ \exp{(\alpha (\mathrm{T}_{\mathrm{obs}}-2003))}
\end{equation}

where $\alpha= -1.67 \times 10^{-3} \  \mathrm{yr}^{-1}$ is the synchrotron fading and $\nu$ and $\mathrm{T}_{\mathrm{obs}}$
are the frequency and the date of observation in years A.C., respectively. \\

From above, we can conclude that the millimetre emission of the Crab nebula has the same physical origin 
than the radio synchrotron emission and therefore it is expected to be polarised with the same degree of polarisation 
and the same orientation. From the current data set we estimate that the errors on the reconstruction of the degree and angle
of polarisation on the Planck cosmological channels induced by an extra component will be well below 1 \%.
This result strongly supports the choice by the Planck collaboration of the 
Crab nebula emission for performing polarisation cross-checks in the range 30 to 353 GHz \cite{Aumont2009}.

\acknowledgments
  We thank the Archeops collaboration for their efforts throughout the
  long campaigns. We acknowledge R. Bandiera and R.D. Davies for very helpful discussions. 
  We finally thank Claudine Tur (LPSC) for fruitful help on bibliographic searches.


\newpage
\begin{thebibliography}{}

\bibitem[Aller \& Reynolds 1985]{Aller1985}
Aller H.~D.~ \& ~Reynolds S.~P., 1985, \apj, 293, L73

\bibitem[Allen \& Barrett 1967]{Allen1967}
Allen R.~J.~\&~Barrett A.~H., 1967, \apj, 149, 1

\bibitem[Aumont 2009]{Aumont2009}
Aumont, J., 2009, \aap, submitted

\bibitem[Baars \& Hartsuijker 1972]{Baars1972}
Baars J.~W.~M., Hartsuijker A.~P.~, 1972, \aap, 17, 172

%
\bibitem[Baars {\it et al.} 1977]{Baars1977}
Baars J.~W.~M. {\it et al.}, 1977, \aap, 61, 99 


\bibitem[Bandiera {\it et al.}  2002]{Bandiera2002}
Bandiera R. {\it et al.}, 2002, \aap, 386, 1044

\bibitem[Becklin \& Kleinmann  1968]{Becklin1968}
Becklin E.~E.~ \& Kleinmann D.~E., 1968, \apj, 152, L25


\bibitem[Chini {\it et al.}   1984]{Chini1984}
Chini R. {\it et al.}, 1984, \aap, 137, 117-127



\bibitem[Dmitrenko {\it et al.}   1970]{Dmitrenko1970}
Dmitrenko T. {\it et al.}, 1970, Radiofizika, Vol. 13, No. 6, 823-829



\bibitem[Desert {\it et al.}   2008]{Desert2007}
Desert F.-X., Mac\'{\i}as-P\'erez J.~F.~, Mayet F. {\it et al.}, 2008, \aap, 481, 411-421 


\bibitem[Flett \& Murray 1991]{1991MNRAS.249.4P}
Flett, A.~M. \& Murray, A.G. 1991, \mnras, 249, 4P


\bibitem[Grasdalen 1979]{Grasdalen}
Grasdalen G.~L., 1979, PACS,  91, 436 

\bibitem[Greaves {\it et~al.} 2003]{2003MNRAS.340..353G}
Greaves, J.~S., Holland, W.~S., Jenness, T., {\it et~al.} 2003, \mnras, 340,
  353

\bibitem[Hinshaw {\it et al.} 2009]{Hinshaw2009}
Hinshaw, B., {\it et al.}, 2009, \apj, 180, 225-245

\bibitem[Green {\it et al.} 2004]{Green2004}
Green D.A., Tuffs R.J. \& Popescu C.C., 2004, \mnras,  355, 1315-1326 

\bibitem[Hobbs {\it et al.}   1968]{Hobbs1968}
Hobbs R.~W., Corbett H.~H., Santini N.~J., 1968, \apj, 152, 43 

\bibitem[Janssen {\it et al.}   1974]{Janssen1974}
Janssen M.~A., Golden L.~M. and Welch W.~J., 1974, \aap, 33, 373-377 

\bibitem[Kalaghan \&  Wulfsberg   1967]{Kalaghan1967}
Kalaghan P.~M. and Wulfsberg K.~N., 1967, Astronomical J. , 72, 1051 

\bibitem[Kellermann {\it et al.}   1969]{Kellermann1969}
Kellermann K.~I., Pauliny-Toth I.~I.~K., Williams P.~J.~S., 1969, \apj, 157, 1
 
\bibitem[Kovalenko {\it et al.}   1994]{Kovalenko1994}
Kovalenko A.~V., Pynzar' A.~V., Udal'tsov V.~A., 1994, Astronomy Reports, 38, 78-94 

\bibitem[Mac\'{\i}as-P\'erez {\it et al.}   2007]{Macias-Perez2007}
Mac\'{\i}as-P\'erez J.~F. {\it et al.}, 2007, \aap, 467, 1313


\bibitem[Marsden {\it et al.}  1984]{Marsden1984}
Marsden P.~L. {\it et al.}, 1984, \apj, 278, L29


\bibitem[Medd     1972]{Medd1972}
Medd W.~J., 1972, \apj, 171, 41


\bibitem[Mezger  {\it et al.}  1986]{Mezger1986}
Mezger P.~G. {\it et al.}, 1986, \aap, 167, 145

\bibitem[Page {\it et al.} 2007]{Page2006}
Page, L. {\it et al.}, 2007, \apj, 170, 335


\bibitem[Penzias \&  Wilson   1965]{Penzias1965}
Penzias A.~A. and Wilson R.~W., 1965, \apj , 142, 1149


\bibitem[Reynolds \& Chevalier 1984]{Reynolds1984}
Reynolds S.~P., Chevalier R.~A., 1984, \apj, 278, 630


\bibitem[Strom \& Greidanus  1992]{Strom1992}
Strom R.~G. \& Greidanus H., 1992, Nature, 358, 654

\bibitem[Veron-Cetty \& Woltjer  1993]{Veron1993}
V\'eron-Cetty M.~P. \& Woltjer L., 1993, \aap, 270, 370

\bibitem[Vinogradova {\it et al.}   1971]{Vinogradova1971}
Vinogradova L.~V.~ {\it et al.}, 1971, Radiofizika, Vol. 14, No. 1, 157-159



\bibitem[Vinyajkin  2005]{Vinyajkin2005}
Vinyajkin E.~N., 2005,  astro-ph/0502033


\bibitem[Werner {\it et al.}  1977]{Werner1977}
Werner M.~W. {\it et al.}, 1977, PACS, 89, 127 


\bibitem[Wright {\it et al.}  1979]{Wright1979}
Wright E.~L. {\it et al.}, 1979, Nature, 279,703 

\bibitem[Wrixon  {\it et al.}  1972]{Wrixon1972}
Wrixon G.~T. {\it et al.}, 1972, \apj , 174, 399

\bibitem[Zabolotnyi {\it et al.}   1976]{Zabolotnyi1976}
Zabolotnyi V.~F., Kostenko V.~I., Slysh V.~I.
1976, Soviet Astronomy,  19, 405  




\end{thebibliography}
\end{document}